# BUILDING TERRESTRIAL PLANETS

A. Morbidelli[1], J.I. Lunine[2], D.P. O'Brien[3], S.N. Raymond[4], K. J. Walsh[5]

[1]Dep. Cassiopée, Universite' de Nice-Sophia Antipolis, Observatoire de la Côte d'Azur, CNRS, B.P.4229, 06304 Nice Cedex 4, France; morby@oca.eu

[2]Department of Astronomy, Cornell University, Ithaca, NY 14853, USA; jlunine@astro.cornell.edu

[3]Planetary Science Institute, 1700 E. Fort Lowell, Tucson, AZ 85719, USA; obrien@psi.edu

[4]Université de Bordeaux, Observatoire Aquitain des Sciences de l'Univers, CNRS, 2 rue de l'Observatoire, BP 89, F-33271 Floirac Cedex, France;rayray.sean@gmail.com

[5]Department of Space Studies, Southwest Research Institute, 1050 Walnut St. Suite 400, Boulder, CO 80302, USA; kwalsh@boulder.swri.edu

**corresponding author:** Alessandro Morbidelli

Dep. Cassiopée, Universite' de Nice-Sophia Antipolis, Observatoire de la Côte d'Azur, CNRS

B.P.4229, 06304 Nice Cedex 4, France;

EMAIL: morby@oca.eu; TEL:(+33) 4 92 00 30 51

**Table of Contents:**

**Abstract**

This paper reviews our current understanding of terrestrial planets formation. The focus is on computer simulations of the dynamical aspects of the accretion process. Throughout the chapter, we combine the results of these theoretical models with geochemical, cosmochemical and chronological constraints, in order to outline a comprehensive scenario of the early evolution of our Solar System. Given that the giant planets formed first in the protoplanetary disk, we stress the sensitive dependence of the terrestrial planet accretion process on the orbital architecture of the giant planets and on their evolution. This suggests a great diversity among the terrestrial planets populations in extrasolar systems. Issues such as the cause for the different masses and accretion timescales between Mars and the Earth and the origin of water (and other volatiles) on our planet are discussed at depth.

1. Introduction

Although we have more information about the terrestrial planets than about virtually any other celestial bodies, the processes leading to their formation have long remained elusive. Only in the last few decades has there been significant progress. On one hand, geochemical and cosmo-chemical analyses performed with laboratory instruments of unprecedented precision have produced a huge amount of data on the chemical and isotopic composition of the planets and of their precursors -- meteorites -- as well as constraints on the chronology of their accretion and thermal evolution. On the other hand, the remarkable increase in computer performance has allowed modelers to undertake increasingly realistic simulations of the dynamical process of terrestrial planet accretion. The results achieved on each front have reached a sufficient level of reliability to be integrated in a comprehensive view of the early evolution of the inner solar system. This review paper will focus on the key processes as seen from an astrophysical point of view but we will make reference to geochemical, cosmochemical and astronomical constraints. Our aim is that this chapter be useful to both theorists and observers in confronting the results of the modeling with laboratory and observational constraints.

Terrestrial planet formation occurred through three distinct modes of growth that were ordered in time in the protoplanetary disk. An observer privileged to watch this process live would see a very distinct distribution of gas and solids, of particles sizes, and of orbits of solid bodies, in each of the three steps; they are potentially distinguishable from each other through disk observations by the next generation of powerful ground and space-based observatories. In step I, planetesimals are formed in a disk of gas and dust. In step II, the collisional evolution of the planetesimal population leads to the growth of a new class of objects called planetary embryos, which represent an intermediate stage between planetesimals and planets. Giant planets probably form at roughly the same time as planetary embryos. In step III, after the disappearance of gas from the proto-planetary disk, the embryos' orbits become unstable, and their mutual collisions give birth to a small number of massive objects, the terrestrial planets. These steps are described in Sects. 2-4. Section 5 details how the results of the simulations change depending on various assumptions (eg. the mass distribution in the disk, giant planet orbits, outcome of collisions etc.). Section 6 discusses a new model that aims to link in a coherent scenario the dynamical evolution of the giant planets with the formation of the terrestrial planets, with the specific goal of explaining the small mass of Mars relative to the Earth. Section 7 discusses the origin of water on Earth and other chemical implications of the terrestrial planet accretion models, while Section 8 details future observing capabilities that might test these ideas.

We dedicate this review to G.W. Wetherill, who was the first to investigate terrestrial planet formation by combining dynamical simulations and geo/cosmo-chemical constraints. He traced the path that we are continuing to follow here.

**2. Step I: from dust to planetesimals.**

When a molecular cloud collapses under its own gravity to form a star, the material forms a disk-like structure in orbit around the central object, due to the conservation of angular momentum. These proto-planetary disks are now routinely observed around pre-main sequence stars (e.g., McCaughrean & O'Dell 1996, Kenyon & Hartmann 1995), while that for the Sun -- the "solar nebula" -- is understood from the physical and chemical evidence left behind in our solar system.

In proto-planetary disks, dust grains sediment into a thin layer at the mid-plane of the disks (Wedenschilling 1980). The sedimentation timescale and the volume density of grains near the mid-plane depend on the severity of turbulence in the disk gas, as strong turbulence inhibits sedimentation. Even in laminar disks, though, the volume density of grains in the mid-plane could not reach arbitrarily large values, because the sedimentation of an excessive amount of solids would itself generate turbulence in the disk via the so-called Kelvin-Helmoltz instability (Weidenschilling 1995).

The growth from these grains to kilometer-size planetesimals is still quite a mystery. In principle, one could expect that grains, once sufficiently concentrated near the mid-plane, should stick to each other to form progressively larger objects, in an ordered-growth process. But particles of cm-size are too small for gravity to be effective in particle-particle collisions, and too big to stick together through electrostatic forces. Moreover, grains are subject to gas drag, which makes them drift towards the central star (Weideschilling 1977). The drift speed is size-dependent; thus, particles of different sizes should collide with non-negligible relative velocities of the order of several cm/s. At these speeds particles should break, rather than coagulate (but see Wettlaufer 2010). Because the drift speed towards the central star is maximal for meter-size boulders, this issue is known as the "meter-size barrier problem", even though it is likely that the fragmentation bottle-neck for accretion starts at much smaller sizes (cm or dm).

A new alternative to this ordered-growth process is that planetesimals form due to the collective gravity of massive swarms of small particles, concentrated at some locations (local maxima of the gas density distribution or inter-vortex regions, depending on particle sizes) by the turbulence of the disk (Johansen et al. 2007; Cuzzi et al. 2008). These "gravoturbulent" models can explain the formation of planetesimals of size 100 km or larger without passing through intermediate small sizes, thereby circumventing the meter-size barrier problem. The size distribution of objects in the asteroid belt and in the Kuiper belt, where most of the mass is concentrated in 100km objects, supports this scenario (Morbidelli et al. 2009; but see Weidenschilling 2010). The existence and the properties of binary Kuiper belt objects also are best explained by the gravitational collapse of massive swarms of small particles that have angular momenta too large to form single objects (Nesvorny et al. 2010). Although more work is needed to explore all the facets of this novel view of planetesimal formation, it seems to resolve many of the outstanding problems that have plagued other models.

In the gravitoturbulent models, once enough small particles are concentrated at some location, the formation of a planetesimal is extremely rapid (Johansen et al. 2007; Cuzzi et al. 2008). However, the formation of self-gravitating clumps of small particles is sporadic (Cuzzi et al. 2010; Chambers 2010), and hence planetesimal formation can in principle proceed over an extended period of time. Moreover, one should not assume that planetesimal formation starts at the same "time zero" in every region of the proto-planetary disk, where "time zero" is usually identified with the formation time of the first solids, the calcium-aluminium inclusions (CAIs, dated at 4.568 Gy ago; Bouvier et al. 2007; Burkhardt et al. 2008). Sufficient clumping of small particles to form planetesimals is possible only if the solid/gas density ratio is larger than some threshold value (Johansen et al. 2009). This ratio in principle increases with time due to the progressive removal of gas from the disk (Throop & Bally 2005). Thus, in some parts of the disk, for instance the innermost regions, this condition might have been met very early, thus leading to a first generation of planetesimals rich enough in short-lived radionuclides (principally $^{26}$Al and $^{60}$Fe; Ghosh et al 2006) to undergo melting or metamorphism of the rocky component.  Farther out in the disk, this condition might have generally been met late enough that the remaining quantities of short-lived radionuclides was too small to allow thermally-driven melting or metamorphism of 100-km scale planetesimals. The threshold for differentiation of 100-km planetesimals might have been reached inward of (Bottke et al. 2006), or even within (Ghost et al. 2006), the asteroid belt.

In an idealized chemical model, planetesimals should accrete all elements and molecules that are in condensed form at their corresponding location of the disk. Since at a given time the temperature in the disk decreases with distance from the Sun, a classic condensation sequence characterized by a clear radial gradient of chemical properties should be obtained (e.g., Dodson-Robinson et al 2009). However, in gravoturbulent models, planetesimals form sporadically so that at any given location they need not have formed at the same time. This effect probably makes the radial gradient of chemical compositions in the planetesimal disk less "clean" than one would predict from a purely condensation-driven sequence.

Observational constraints are available to deduce properties and confront models of the planetesimal disk. There are three classes of chondritic meteorites: enstatite, ordinary and carbonaceous. Their chemistry and mineralogy suggest that these three classes formed respectively at decreasing temperatures. For instance, water is essentially absent on enstatite meteorites, and quite abundant in (some subclasses of) carbonaceous chondrites, while the water-content in ordinary chondrites is intermediate between the two (Robert 2003). Spectroscopic observations link these three classes of meteorites to asteroids of different taxonomic type (although not necessarily to specific parent bodies): enstatite chondrites can be linked with E-type asteroids (Fornasier et al. 2008), which are predominant in the Hungaria region at 1.8 Astronomical units (AU; 1 AU is the mean Sun-Earth distance); ordinary chondrites are linked to S-type asteroids (Binzel et al. 1996), which are predominant in the inner belt (2.1-2.8 AU); and carbonaceous chondrites are linked to C-type asteroids (Burbine 2000), predominant in the outer belt (beyond 2.8 AU). There is, however, substantial overlap in the radial distribution of these three types of asteroids (Gradie and Tedesco 1982). This could be the result of the formation of subsequent generations of asteroids in a cooling disk, as mentioned before. However, the analysis of the ages of the individual chondrules in meteorites does not show any clear difference in accretion age of the parent bodies of ordinary and carbonaceous chondrites (Villeneuve et al. 2009), which both seem to have formed 3-4 My after CAI formation. This supports the alternative idea that the partial overlapping of the radial distribution of asteroids of different types is due to dynamical mixing that occurred after the formation of the asteroids (see sections 4 and 6).

Comets are representative of the planetesimal disk that formed at larger distances than the asteroid belt, i.e. in between and beyond the giant planets orbits. The classical view is that, while the parent bodies of carbonaceous chondrites are rich in hydrated minerals, comets are rich in water ice and lack hydrated silicates, presumably because they formed in a colder environment. However, new discoveries are making the difference between carbonaceous chondrites (or C-type asteroids) and comets less well defined. The close flyby images of comets (e.g. Comet Borrelly) show very little surface ice and small active regions (Sunshine et al. 2006). The samples from comet Wild 2 by the Stardust Mission turned out to be quite similar to meteoritic samples (Zolensky et al. 2006). Modeling work on the origin of the dust that produces the zodiacal light (Nesvorny et al. 2010b) predicts that at least 50% of the micro-meteorites collected on Earth are cometary; however, we see no clear separation of micro-meteorites into two categories, which could be traced to asteroidal and cometary dust (Levison et al. 2009). Water-ice has been found on the C-type asteroid Themis (Campins et al. 2010; Rivkin et al. 2010) and some C-type asteroids in the main belt show cometary activity (Hsieh and Jewitt 2006). The possibility of a continuum in physical and chemical properties between carbonaceous asteroids and comets is well described in Gounelle et al. (2008).

Putting all this information together in a coherent picture is not a simple task. But at the very least we can say that there is evidence that a radial gradient in temperature existed in the disk at the time(s) when planetesimals formed. In particular, planetesimals in the inner disk (in the inner asteroid belt region and presumably also in the terrestrial planet region) appear to have been dry and volatile poor. As pointed out by Albarede (2009), the gaseous Solar Nebula probably dissipated before the temperature decreased enough to allow volatiles to condense in the inner solar system, thus explaining the lack of close-in volatile-rich primitive objects (i.e. objects with solar composition). This consideration opens the question of how water and other volatiles have been delivered to the Earth, which we will address in section 7.

## 3. Step II: from planetesimals to planetary embryos

Once the proto-planetary disk contains a substantial population of planetesimals and the dynamics of accretion becomes dominated by the effect of the gravitational attraction between pairs of planetesimals, the second stage of planet formation can start. During this phase of accretion -- *runaway growth*, -- big bodies grow faster than the small ones and hence rapidly increase their relative mass difference (Greenberg et al 1978). This process can be formalized by the inequality:

$$d/dt\,(M_1/M_2) > 0,$$

where $M_1$ and $M_2$ are respectively the characteristic masses of the "big" and of the "small" bodies. The reasons for runaway growth can be explained as follows.

At the beginning of the runaway growth phase the largest planetesimals, by necessity, must represent only a small fraction of the total mass. Hence the dynamics is governed by the small bodies, in the sense that the relative velocities between bodies is of order of the escape velocity of the small bodies $V_{esc(2)}$. This velocity is independent of the mass $M_1$ of the large bodies and is smaller than their escape velocity $V_{esc(1)}$. For a given body, its collisional cross-section is enhanced with respect to the geometrical cross-section by the so-called *gravitational focusing* factor $F_g$, so that :

$$dM/dt \sim R^2\,F_g$$

The gravitational focusing factor is given by:

$$F_g = 1 + V_{esc}^2/V_{rel}^2$$

(Safronov & Zvjagina 1969, Greenberg et al 1978, Greenzweig and Lissauer 1992), where $V_{esc}$ is the body's escape velocity and $V_{rel}$ is the relative velocity of the other particles in its environment. Because $V_{rel} \sim V_{esc(2)}$, the gravitational focusing factor of the small bodies (with $V_{esc}=V_{esc(2)}$) is of order unity, while that of the large bodies (with $V_{esc}=V_{esc(1)}>>V_{esc(2)}$) is much larger, of order $V_{esc(1)}^2/V_{rel}^2$. In this situation, since both $V_{esc(1)}$ and the geometrical cross section are proportional to $M_1^{2/3}$, the mass growth of a big body is described by the equation

$$1/M_1 \, dM_1/dt \sim M_1^{1/3} V_{rel}^{-2}$$

(Ida and Makino 1993). Therefore, the relative growth rate is an increasing function of the body's mass, which is the condition for the runaway growth (Fig.1).

Runaway growth stops when the masses of the large bodies become important (Ida and Makino 1993) and the perturbations from the large bodies begin to govern the dynamics. The condition for this to occur is:

$$n_1 M_1^2 > n_2 M_2^2,$$

where $n_1$ (resp. $n_2$) is the number of big bodies (resp. small bodies). In this case, $V_{rel} \sim V_{esc(1)}$, so that $F_g \sim 1$; hence $(1/M_1)(dM_1/dt) \sim M_1^{-1/3}$. The growth rate of the embryos gets slower as the bodies increase in size and the relative differences in mass among the embryos also slowly decreases. In principle, one might expect that the small bodies continue to grow, approaching the mass of the embryos, but in reality, the now large relative velocities prevent the small bodies from accreting each other: their mutual collisions rather become disruptive. Thus, the small bodies can only participate in the growth of the embryos. This phase is called `oligarchic growth' (Kokubo & Ida 1998, 2000, Chambers 2006).

The runaway growth phase happens throughout the disk on a timescale that depends on the local dynamical time (keplerian time), on the planetesimal size and on the local density of available solid material. This density will also determine the maximum size of the embryos when the runaway growth ends (Lissauer 1987). Assuming a reasonable surface density of solid materials, the runaway growth process forms planetary embryos of lunar to martian mass at 1AU in $10^5$-$10^6$y, separated from each other by a few times $10^{-2}$ AU.

Thus, the planetary embryos are not yet the final terrestrial planets (at least Earth and Venus). They are not massive enough, they are too numerous and they are too closely packed relative to the terrestrial planets that we know. Moreover they form too quickly, compared to the timescale of several $10^7$y suggested for the Earth by radioactive chronometers (Kleine et al. 2009).

Because runaway growth is a local process, the embryos form from the planetesimals in their neighborhoods. Little radial mixing is expected at this stage. Thus, if the planetesimal disk is characterized by a radial gradient of chemical properties, such a gradient should be reflected in the embryos' chemical compositions.

In view of their large mass and their rapid formation timescale, embryos can undergo differentiation. We stress, though, that embryo formation cannot be faster than planetesimal formation, because a massive planetesimal population is needed to trigger the runaway growth of the embryos. Thus, if planetesimals in the asteroid belt accreted about 3-4 My after CAIs (Villeneuve et al. 2009; see description of step 1), the embryos in the asteroid belt cannot have formed earlier than this time. In this case, the low content of short-lived radioactive elements might have allowed them to avoid differentiation, similarly to what is invoked for Callisto (Canup & Ward 2002) or Titan (Sotin et al. 2010). The lack of differentiation could have helped the embryos formed in the outer asteroid belt to preserve the water inherited from the local carbonaceous chondrite-like planetesimals. Nevertheless, even if differentiation had occurred on water-rich embryos, water would not have been necessarily lost; water ice could have formed a mantle around a rocky interior, possibly differentiated itself into a metallic core and a silicate outer layer, like on Europa and Ganymed.

The formation of the giant planets is intimately related to the runaway/oligarchic growth of embryos. Beyond the so-called *snow line* between 2 - 4 AU, the low temperatures allow water ice to condense, enhancing the surface density of solid material and drastically increasing the typical embryo mass to perhaps an Earth mass or more (e.g., Stevenson & Lunine 1988, Kokubo & Ida 2002, Ciesla & Cuzzi 2006, Dodson-Robinson et al 2009). The formation of the massive cores of the giant planets (of about 10 Earth masses each) remains poorly understood. It has been proposed that convergent migration processes could have brought these embryos together, favoring their rapid mutual accretion (Morbidelli et al. 2008; Lyra et al. 2009; Sandor et al. 2011). Once formed, the cores started to accrete massive atmospheres of hydrogen and helium from the proto-planetary disk, thus becoming the giant planets that we know. The formation of gas giant planets is a complex subject that merits its own review (see, for example, Lissauer & Stevenson 2007). What is most important for our purposes is that giant planets form quickly, before the final step of terrestrial planet formation, and therefore influence this last phase of terrestrial accretion.

**4. Step III: from embryos to terrestrial planets: overview**

At the time of the disappearance of the gas from the proto-planetary disk, the solar system should have had the following structure: (i) in the inner part, a disk of planetesimals and planetary embryos with roughly equal total mass in each component; (ii) in the central part, a fully-formed system of giant planets; and (iii) beyond the orbits of the giant planets, another disk of planetesimals. The orbits of the giant planets were inherited from their previous dynamical evolution dominated by their gravitational interactions with the gas-disk, and they were likely different from the current orbits. We will come back to this important issue in the following sections.

When the nebular gas is present, it has a stabilizing effect on the system of embryos and planetesimals because it continuously damps their orbital eccentricities. Thus, when the gas is removed, the eccentricities can grow rapidly, leading to intersecting orbits and collisions among embryos (Chambers and Wetherill 1998). However, numerical simulations (Chambers and Wetherill 2001; Raymond et al. 2004, 2005; O'Brien et al. 2006) show that the dynamical evolution is very different in the terrestrial planet region versus the asteroid belt. In the terrestrial planet region, where the perturbations exerted by Jupiter are weak, the embryos' eccentricities remain relatively small, and the embryos can accrete each other in low velocity collisions. However, the asteroid belt is crossed by several powerful resonances with Jupiter that excite the eccentricities of the resonant objects. The orbits of the embryos and planetesimals are continually changing due to mutual encounters; additionally, every time that one of them temporarily falls into a resonance, its eccentricity is rapidly enhanced. As a result, most of the original population of embryos and planetesimals eventually leaves the asteroid belt region by acquiring orbits that are eccentric enough to collide with the Sun or cross Jupiter's orbit and be ejected on a hyperbolic orbit; a fraction of them can be accreted by the growing planets inside of 2 AU, and a small number of planetesimals remain on stable orbits in the belt (see Fig. 2 for an illustration of this process). The typical result of this highly chaotic phase --simulated with several numerical N-body integrations-- is the elimination of all the embryos originally situated in the asteroid belt and the formation of a small number of terrestrial planets up to of order the mass of the Earth on stable orbits in the 0.5--2 AU region in a timescale of several tens of Myr (eg. Chambers and Wetherill 2001; Raymond et al. 2004, 2005; O'Brien et al. 2006).

This scenario has several strong points:

(i) Typically, 2 to 4 planets are formed on well-separated and stable orbits. If the initial disk of embryos and planetesimals contains about 4-5 Earth masses of solid material, typically the two largest planets are about one Earth mass each. Moreover, in the most modern simulations, accounting for the dynamical interaction between embryos and planetesimals (O'Brien et al. 2006; see also Morishima et al 2010, Raymond et al 2009), the final eccentricities and inclinations of the terrestrial planets that form are comparable or even smaller than those of the real planets (an explanation of this fact is provided in Sect. 5.3).

(ii) Quasi-tangent collisions of Mars-mass embryos onto the proto-planets are quite frequent (Agnor et al. 1999; Morishima et al. 2010, Elser et al 2011). These collisions are expected to generate a disk of ejecta around the proto-planets (Canup & Asphaug 2002), from which a satellite is likely to accrete (Canup and Esposito 1996). This is the standard, generally accepted, scenario for the formation of the Moon (e.g., Benz et al. 1987).

(iii) The accretion timescale of the terrestrial planets in the simulations is ~30-100 Myr, in general agreement with the timescale of Earth accretion deduced from radioactive chronometers (whose estimates vary from one study to another over a comparable range; Yin et al. 2002; Touboul et al., 2007; Allègre et al. 1995).

(iv) A small fraction of the original planetesimals typically remain in the asteroid belt on stable orbits at the end of the terrestrial planet formation process and all embryos are ejected from the belt in most, but not all, simulations (Petit et al 2001; O'Brien et al. 2007). The depletion of the belt by embryos and their subsequent removal is essential to explaining the asteroid belt as we see it today, including its substantial mass deficit. The orbital eccentricities and inclinations of these surviving particles compare relatively well to those of the largest asteroids in the current belt. Moreover, because of the scattering suffered from the embryos, the surviving particles are randomly displaced in semi major axis, relative to their original position, by about 0.5 AU. This can explain the partial mixing of asteroids of different taxonomic types, discussed in section 2.

However, not all simulations are equally successful. A discussion on how the results change depending on parameters and initial conditions is reported in the next section. Moreover, a general problem of the simulations is that the synthetic planet produced at the approximate location of Mars is systematically too massive (Wetherill 1991, Chambers 2001, Raymond et al. 2009). Mars is an oddity not only for what it concerns its mass, but also its accretion timescale: in fact, it formed in a few millions of years only, like asteroids, i.e. much faster than the Earth (Dauphas & Pourmand 2011). We discuss these issues in more detail in section 6.

## 5. From embryos to terrestrial planets: Dependence of the results on simulation and model parameters

*5.1 Outcome of giant collisions*

In all simulations of the accretion of terrestrial planets (with the exception of Alexander & Agnor 1998 and Kokubo and Genda 2010 - see below) it has been assumed that all collisions between embryos are perfectly accretional, i.e. every time two embryos collide, they merge. However, this is a gross simplification. Simulations of giant collisions between embryos conducted with the Smooth Particle Hydrodynamics (SPH) technique (Agnor and Asphaug 2004; Asphaug et al. 2006) show that perfect merging is rare. In some cases most of the masses of the two embryos merge, but a fraction of the total mass is ejected into space in the form of small objects, or fragments. In other cases, there are "hit and run collisions" where the embryos do not merge: they bounce off each other, again losing part of their masses as ejecta. The consequences of "imperfect collisions" are not fully understood, although preliminary studies have given us some clues.

Kokubo and Genda (2010) investigated the effects of bouncing collisions on the dynamics of terrestrial planet accretion. From a database of SPH simulations of embryo-embryo collisions, they first derived accretion conditions in terms of impact velocity, angle and masses of colliding bodies. Then, in N-body simulations of terrestrial planet accretion, they treated each collision either as a perfect merger or an inelastic bounce as determined by comparing the impact parameters of the collision with comparable cases in the collision database. They found that, on average, half of collisions do not lead to accretion. However, the final number, mass, orbital elements, and even growth timescale of the terrestrial planets are barely affected by this large fraction of bouncing collisions. The reason is that, if two embryos bounce off each other, they both remain in the system and tend to accrete during their next encounter. Thus, having accretion at the first collision or, on average, every other collision, has very little influence on the broad evolution of the simulation.

In reality, of course, collisions are neither perfect mergers nor bounces. As SPH simulations show, part of the colliding bodies' mass is merged and part is released to space as collisional fragments. These fragments might be quickly re-accreted by the embryo, leading to little net change in the overall planet formation history, as found by Kokubo and Genda (2010). Alternately, the fragments could undergo collisional grinding into dust and be removed by radiation forces, meaning that a fraction of the total disk mass would be lost.

The interest in the latter possibility lies in the possible consequences that it can have on the chemistry of the terrestrial planets. If most planetary embryos are differentiated, the material that is released to space during giant collisions should come preferentially from the mantles of the embryos rather than their cores. This would lead to a change in the bulk composition of the growing terrestrial planets relative to the original composition of their progenitors, particularly increasing the final iron/silicate ratio (O'Neill & Palme 2008). The metal content of the terrestrial planets is indeed higher than in any undifferentiated meteorite (Jarosewich 1990).

*5.2 Disk mass and radial profile*

Kokubo et al. (2006) performed the most recent systematic exploration on how the outcomes of the terrestrial planet formation process depend on the basic properties of the proto-planetary disk (surface density, surface density profile, orbital separation of the initial protoplanet system, and bulk density of protoplanets). For computational reasons, in all simulations, the disk of solids was assumed to range from 0.5 to 1.5 AU. No giant planets were considered. For the standard disk model, with a surface density of solids equal to $10 g/cm^2$ at 1 AU and decaying as $1/r^{3/2}$, typically two Earth-sized planets formed in the terrestrial planet region. The number of planets slowly decreased as the surface density of the initial protoplanets was increased, while the masses of individual planets increased almost linearly. The basic structure of the resulting planetary systems depended only slightly on the initial distribution of protoplanets (individual masses and mutual separations in units of Hill radius) and the bulk density, as long as the total mass was fixed. For a steeper surface density profile, large planets formed on average closer to the star. The dependence of the results on the disk's surface density profile had also been investigated in Raymond et al. (2005), who found that, for steeper profiles the final planets were more numerous, formed more quickly, were more massive, and had higher iron contents and lower water contents.

*5.3 Particle size distribution in the disk*

The first simulations of the terrestrial planet accretion process using modern integration techniques (Chambers and Wetherill 1998, 2001; Agnor et al. 1999; Chambers 2001) used as initial conditions a set of a few tens of planetary embryos and produced planets on orbits that were systematically too eccentric and inclined. A commonly-used measure of the orbital excitation of the terrestrial planets is the *normalized angular momentum deficit* (AMD), defined as (Laskar 1997):

$$\text{AMD} = \{\Sigma_j\, m_j\, [a_j(1-e_j^2)]^{1/2} \cos i_j - m_j a_j^{1/2}\} / \{\Sigma_j\, m_j\, a_j^{1/2}\}$$

where the sum is calculated over all the planets, with mass $m_j$, semi major axis $a_j$, eccentricity $e_j$ and inclination $i_j$. For reference, the AMD of the actual terrestrial planets is -0.0018, when averaged over million-year timescales, compared with a median value of -0.0050 in the simulations of Chambers (2001). The small orbital excitation of the real terrestrial planets, therefore, was a major property that the simulations failed to reproduce for many years.

To solve this problem, one proposal was that a small fraction of gas from the proto-solar nebula was still present in the system when the terrestrial planets formed, so that the eccentricities and inclinations of embryos and proto-planets were continuously damped by the gravitational interactions with the gas. However, the simulations that accounted for this process (Kominami and Ida 2002, 2004) systematically resulted in too many terrestrial planets, forming in less than 10 My (too short a timescale, relative to radioactive chronometer constraints; Halliday and Klein 2006), on quasi-circular orbits that are much closer to each other than the real orbits. Ogihara et al. (2007) alleviated the problem of the number and separation of terrestrial planets, by assuming that the disk was strongly turbulent; nevertheless, the problem of the excessively short formation timescale remained (see also Thommes et al. 2008).

A different approach to solving the problem of the small orbital excitation of the terrestrial planets was to explore different mass distributions in the original disk of solid bodies. Chambers (2001) showed that, starting the simulations with a larger number of smaller embryos than in Chambers and Wetherill (1998), resulted in a significant reduction of the final AMD, although the latter was still too large relative to that of the real terrestrial planets. This trend continued in the simulations of Raymond et al. (2004, 2005, 2006b, 2007) and eventually led to AMDs consistent with that of the real system (Raymond et al. 2009; Hansen 2009).

Instead, O'Brien et al. (2006) assumed that the initial disk consisted of a population of planetary embryos, well separated from each other, embedded in a population of planetesimals. The total mass of each population was the same. In total, 25 Mars-mass embryos were considered from 0.3 to 4 AU, while the planetesimal population was modeled with 1,000 equal mass particles. The planetesimals were assumed to interact with the embryos, but not with each-other. With these initial conditions, some simulations (those with Jupiter and Saturn initially on their current orbits, see Sect. 5.4) achieved, for the first time, a system of synthetic terrestrial planets with a AMD = -0.001, even smaller than the real one (the reader should keep in mind that the terrestrial planets might have achieved the current AMD after their formation, during a phase of late orbital migration of the giant planets; Brasser et al. 2009).

The success of these simulations argues strongly that the key to explaining the small AMD of the terrestrial planets is *dynamical friction* (Wetherill and Stewart 1993). The latter is the result of the gravitational interaction between populations of bodies of different individual masses. As a result of equipartition of orbital excitation, more massive bodies tend to remain on less eccentric and inclined orbits than less massive bodies. Thus, planetesimals acquire orbits that are more and more excited (until they are mostly removed by collisions with the Sun or ejection onto hyperbolic orbits) while the eccentricities and inclinations of the embryos and forming planets are progressively damped. Of course, given the prevalence of giant impacts during the late stages of growth, embryos must acquire large enough eccentricities for their orbits to cross. However, the key for maintaining low-eccentricity final orbits is the presence of planetesimals *after* giant impacts; collisional debris could play a role in this process, especially as they are naturally generated by the impacts themselves.

*5.4 Giant planet architecture*

There is a clear dependence of the final outcomes of the simulations on the orbital architecture assumed for the giant planets. By definition, the giant planets must be fully formed by the time the gas is removed from the disk, i.e. well before the formation of the terrestrial planets is complete. Thus, in most terrestrial planet formation simulations, the orbital architecture of the giant planet system is considered as an "environmental property" of the inner solar system, i.e., an initial condition.

Terrestrial planet formation with orbital configurations of giant planets vastly different than our solar system was investigated by Levison and Agnor (2003). They found that the number, mass, and location of the terrestrial planets are directly related to the amount of dynamical excitation experienced by the planetary embryos near 1 AU, which in turn is related to the proximity, mass and eccentricity of the giant planets. In general, if the embryos' eccentricities are large, each is crossing the orbits of a larger fraction of its cohorts, which leads to a smaller number of more massive planets. In addition, embryos tend to collide with objects near their periastron. Thus, in systems where the embryos' eccentricities are large, planets tend to form close to the central star. Comparable results were found by Raymond (2006), who focused on the giant planet configurations that permit habitable terrestrial planets to form.

For systems of giant planets resembling our Solar System (Jupiter- and Saturn-mass planets with moderate orbital eccentricities, comparable to those of their current orbits), the results are more subtle. Unexpectedly, the simulations with larger orbital eccentricities of the giant planets form terrestrial planets on a shorter timescale and on more circular final orbits than those with more circular giant planets. Moreover, for eccentric giant planets the terrestrial planets accrete less material from the asteroid belt and the synthetic planet produced at the orbital location of Mars is smaller (Raymond et al. 2009). All these properties are interrelated. Eccentric giant planets deplete the asteroid belt more rapidly compared to giant planets on more circular orbits, so that embryos and planetesimals originally in the belt are removed by collisions with the Sun or ejection on hyperbolic orbit before they have a chance to interact with the growing terrestrial planets inside of 2 AU. Thus, in the presence of eccentric giant planets, terrestrial planet formation proceeds as in a "closed" system, with little or no material from outside of 1.5-2.0 AU incorporated in surviving planets. So, accretion proceeds faster because the material that builds the planets is more radially confined. On the other hand, faster accretion timescales imply that more planetesimals are still in the system at the end of the terrestrial planet accretion process, so that the orbital eccentricities of the terrestrial planets can be damped more by dynamical friction. Finally, Mars forms smaller because it is close to the edge of the radial distribution of the mass that participates in the construction of the terrestrial planets.

Given that the giant planets probably underwent planetesimal-driven orbital migration after the dissipation of the gaseous protoplanetary disk (Fernandez & Ip 1984, Malhotra 1995, Tsiganis et al 2005), the orbits of Jupiter and Saturn at the time of terrestrial planet formation are unknown. To understand which giant planet orbital architecture was more likely at early times, one has to turn attention to the dynamical evolution that the giant planets should have had in the gaseous Solar Nebula. It is well known that, by interacting gravitationally with the gaseous disk, the orbits of the giant planets migrate (see Ward 1997, for a review; note that this process is distinct from "planetesimal-driven" migration). Eventually the planets tend to achieve a multi-resonance configuration, in which the period of each object is in integer ratio with that of its neighbor (Morbidelli et al. 2007). The interaction with the disk also damps the planets' orbital eccentricities. Thus, at the disappearance of the gas disk, the giant planets should have been closer to each other, on resonant and quasi-circular orbits (the giant planets could have achieved their current orbits at a much later time, corresponding to the so-called Late Heavy Bombardment; see Morbidelli 2010 for a review). However, simulations show that this kind of orbital configuration of the giant planets systematically leads to synthetic planets at ~1.5 AU that are much more massive than the real Mars (Raymond et al. 2009).

## 6. The Grand Tack Scenario

Hansen (2009) convincingly showed that the key parameter for obtaining a small Mars is the radial distribution of the solid material in the disk. If the outer edge of the disk of embryos and planetesimals is at about 1 AU, with no solid material outside of this distance, even simulations with giant planets on circular orbits systematically produce a small Mars (together with a big Earth). The issue is then how to justify the existence of such an outer edge and how to explain its compatibility with the existence of the asteroid belt between 2 and 4 AU. The asteroid belt has today a very little total mass (about $6\times10^{-4}$ Earth masses; Krasinsky et al. 2002), but it must have contained at least a thousand times more solid material when the asteroids formed (Wetherill 1989).

The result by Hansen motivated Walsh et al. (2011) to look in more details at the possible orbital history of the giant planets and their ability to sculpt the disk in the inner solar system. For the first time, the giant planets were not assumed to be on static orbits (even if different from the current ones); instead Walsh et al. studied the co-evolution of the orbits of the giant planets and of the planetesimal and embryo precursors of the terrestrial planets, during the era of the disk of gas. Walsh et al. built their model on previous hydro-dynamical simulations showing that the migration of Jupiter can be in two regimes: when Jupiter is the only giant planet in the disk, it migrates inwards (Lin and Papaloizou 1986), but when paired with Saturn both planets typically migrate outward, locked in a 2:3 mean motion resonance (where the orbital period of Saturn is 3/2 of that of Jupiter; Masset & Snellgrove 2001; Morbidelli & Crida 2007). Thus, assuming that Saturn formed later than Jupiter, Walsh et al. envisioned the following scenario: first, Jupiter migrated inwards while Saturn was still growing; then, when Saturn reached a mass close to its current one, it started to migrate inwards more rapidly than Jupiter, until it captured the latter in the 3/2 resonance (Masset and Snellgrove 2001; Pierens and Nelson 2008); finally the two planets migrated outwards together until the complete disappearance of the disk of gas. The extent of the inward and outward phases of migration are unconstrained a priori, because they depend on properties of the disk and of giant planet accretion that are unknown, such as the time-lag between Jupiter and Saturn's formation, the speed of inward migration (which depends on the disk's viscosity), the speed of outward migration (which depends on the disk's scale height), and the time-lag between the capture in resonance of Jupiter and Saturn and the photo-evaporation of the gas. However, the extent of the inward and outward migration of Jupiter can be deduced by looking at the resulting structure of the inner solar system. In particular, Walsh et al. showed that a reversal of Jupiter's migration at 1.5 AU would provide a natural explanation for the existence of the outer edge at 1 AU of the inner disk of embryos and planetesimals, required to produce a small Mars (see Fig.3,4). Because of the prominent reversal of Jupiter's migration that it assumes, Walsh et al. scenario is nicknamed "Grand Tack".

Several giant extra-solar planets have been discovered orbiting their star at a distance of 1-2 AU, so the idea that Jupiter was at one time 1.5 AU from the Sun is not shocking by itself. A crucial diagnostic of this scenario, though, is the survival of the asteroid belt. Given that Jupiter should have migrated through the asteroid belt region twice, first inwards, then outwards, one could expect that the asteroid belt should now be totally empty. However, the numerical simulations by Walsh et al. show that the asteroid belt is first fully depleted by the passage of the giant planets, but then, while Jupiter leaves the region for the last time, it is re-populated by a small fraction of the planetesimals scattered by the giant planets during their migration. In particular, the inner asteroid belt is repopulated mainly by planetesimals that were originally inside the orbit on which Jupiter formed, while the outer part of the asteroid belt is repopulated mainly by planetesimals originally in between and beyond the orbits of the giant planets. Assuming that Jupiter initially formed at the location of the snow line, it is then tempting to identify the planetesimals originally closer to the Sun with the anhydrous asteroids of E- and S-type and those originally in between and beyond the orbits of the giant planets with the "primitive" C-type asteroids. With this assumption, the Grand Tack scenario can explain the co-existence of asteroids of different types in the main belt physical structure of the asteroid belt in a natural way. In fact, it is difficult to explain the differences between ordinary/enstatite chondrite (E- and S-type) and carbonaceous chondrite (C-type) parent bodies if they had both formed in the asteroid belt region, given that they are coeval (Villeneuve et al. 2009) and that the radial extent of the asteroid belt is small (only ~1 AU). Instead, if ordinary/enstatite and carbonaceous chondrite parent bodies were implanted into the asteroid belt from originally well separated reservoirs, as in the Grand Tack model, the differences in physical properties are easier to understand in the framework of the classical condensation sequence. The origin of C-type asteroids from the giant planet region would also explain, in a natural way, the similarities with comets that are emerging from recent observational results and sample analyses. The small mass of the asteroid belt and its eccentricity and inclination distribution are also well reproduced by the Grand Tack scenario.

This scenario also explains why the accretion timescales of Mars and the asteroids are comparable(Dauphas and Pourmand 2011). In fact, the asteroids stopped accreting when they got dispersed and injected onto excited orbits of the main belt; Mars stopped accreting when the inner disk was truncated at 1 AU and the planet was pushed beyond this edge by an encounter with the proto-Earth (Hansen 2009). In the Grand Tack scenario these two events coincide, and mark the time of the passage of Jupiter through the inner solar system.

All these results on the asteroid belt, together with the fact that the mass distribution of the terrestrial planets is also statistically reproduced (see Fig.4) make the Grand Tack scenario an appealing comprehensive model of terrestrial planet formation.

## 9. Origin of Terrestrial water

Laboratory analyses show a clear correlation between the water content in meteorites and the heliocentric distance of the parent bodies from which said meteorites are thought to come from (See Fig.5). Although there is the theoretical possibility that water-rich planetesimals formed in the hot regions of the disk by water-vapor absorption on silicate grains (Muralidharan et al. 2008; King et al. 2010), the empirical evidence strongly suggests that the planetesimals in the terrestrial planet region were extremely dry.

In contrast, the Earth has a water content that, although small, is non-negligible and definitely larger than what the above-mentioned correlation would suggest for material condensed at 1 AU. In fact, the mass of the water contained in the Earth's crust (including the oceans and the atmosphere) is $2.8 \times 10^{-4}$ Earth masses; the mass of the water in the present-day mantle is estimated to be in the range $0.8-8 \times 10^{-4}$ Earth masses (Lecuyer et al. 1998). Thus, the current bulk water-content on Earth is presumably in the range $5 \times 10^{-4} - 10^{-3}$. However, a larger quantity of water might have resided in the primitive Earth and been subsequently lost during core formation and impacts. Thus, the current Earth has water content larger than enstatite chondrites and it is possible that the primitive Earth had water content comparable to or larger than ordinary chondrites. Where did this water come from, if the local planetesimals were dry?

An important constraint for models on the origin of the Earth's water is the $DHO/H_2O$ ratio. This ratio in the present day mantle and in the oceans is $1.49\pm3\times10^{-4}$ (Lecuyer et al. 1998), which is about 6 times higher than for molecular hydrogen in the proto-planetary disk (deduced from measurements in the atmosphere of Jupiter; Mahaffy et al. 1998). It is believed that the recycling of water in the deep mantle does not significantly change the D/H ratio. However, the D/H ratio could have significantly increased if the Earth had in the past a massive hydrogen atmosphere (with a molar ratio $H_2/H_2O$ larger than 1), that experienced a slow hydro-dynamic escape (Genda and Ikoma 2008).

Presuming that water was not present in sufficient quantities in the local planetesimals and embryos around 1 AU, there are three potential sources for the origin of water on Earth.

The first model invokes a *nebular origin*. Ikoma and Genda (2006) assumed that at the end of the Earth's formation there was still some nebular hydrogen in the proto-planetary disk, and the Earth gravitationally captured a hydrogen-rich atmosphere of nebular gas of up to $10^{21}$ kg in mass. Then, the atmospheric hydrogen was oxidized, perhaps by FeO in the magma ocean, to produce water. However, in this model, the initial D/H ratio of the water would be solar. The D/H ratio can be subsequently increased by hydro-dynamic escape of the H-rich atmosphere (Genda & Ikoma 2008). However, the factor of 6 increase in D/H ratio required to match observations yields unrealistically long timescales, i.e. the hydro-dynamic escape of the primitive atmosphere should have occurred over billions of years, in contrast with constraints from the $^{129}$I/$^{129}$Xe chronometer (Wetherill 1975).

A second possibility is that the water was brought to Earth by the bombardment of comets (Delsemme 1992, 1999). A first problem with this model is that, so far, the observed D/H ratio in the water vapor released by long period comets is about twice that on Earth (Balsiger et al. 1995; Eberhardt et al. 1995; Meier et al. 1998; Bockelee-Morvan et al. 1998). There are no known terrestrial processes that could *decrease* the D/H ratio of the original water. A second problem is that the collision probability of comets with the Earth is very small. Of the planetesimals scattered by the giant planets from the proto-planetary disk, only 1 in a million would strike our planet (Morbidelli et al. 2000). From studies on the range of radial migration that the giant planets should have suffered after the disappearance of the disk of gas, due to their interaction with planetesimals (Malhotra 1995; Hahn and Malhotra 1999; Gomes et al. 2004, 2005), it is expected that the total mass of the cometary disk was 35-50 Earth masses; moreover, from measurements of the ice/dust ratio in comets (Kuppers et al. 2005), it is now believed that less than half of the mass of a comet is in water-ice. Putting all these elements together, the mass of water delivered by comets to the Earth should have only been ~2.5x10$^{-5}$ Earth masses (neglecting impact losses), i.e. only 10% of the crustal water.

The third possibility is that the Earth accreted water from primitive planetesimals and/or planetary embryos originally from the outer asteroid belt (Morbidelli et al, 2000; Raymond et al., 2004, 2005, 2006, 2007, O'Brien et al., 2006; Lunine et al 2007). From an isotopic point of view, this makes sense because the average D/H ratio of water in carbonaceous chondrites is almost identical to that of the Earth (1.59x10$^{-4}$, with individual values ranging over 1.28-1.80x10$^{-4}$; Dauphas et al. 2000). To address the delivery of water by asteroids from the modeling point of view, we need to distinguish between the classical scenario, in which the outer belt is originally inhabited by primitive objects that are removed by mutual scattering and interactions with resonances with Jupiter (Sec. 4), and the Grand Tack scenario (Sec. 6). In the first case, as we have seen above, the amount of material accreted by the terrestrial planets from the asteroid belt depends on the eccentricity of the orbit of Jupiter. If Jupiter's orbit was almost circular, the terrestrial planets should have accreted 10-20% of their mass from beyond 2.5 AU (O'Brien et al. 2006, Raymond et al 2006b), most of which should have been of carbonaceous chondritic nature. Thus, in the classical scenario the terrestrial planets should have originally been very water-rich, possibly even as much as envisioned by Abe et al. (2000), and should have lost most of their water during impacts and/or their geophysical evolution. However, Drake and Righter (2002) have argued that the detailed geochemistry of the carbonaceous chondrites is inconsistent with more than a few percent of the Earth being composed of this material. Furthermore, the amount of material accreted from the outer asteroid belt drops with increasing eccentricity of Jupiter. If Jupiter had had originally an orbit with an eccentricity comparable or larger than the current one, little material would have been accreted from the outer asteroid belt, and the terrestrial planets would have been almost completely dry (O'Brien et al. 2006; Raymond et al. 2009), as noted in section 5.4.

In the Grand Tack scenario, the primitive asteroids (like the C-types) were implanted into the asteroid belt from in between and beyond the orbits of the giant planets. In Walsh et al.'s (2011) simulations, for every primitive planetesimal implanted in the outer asteroid belt, 10–30 planetesimals ended up on orbits that crossed the terrestrial planet forming region, for a total of 3–11x10$^{-2}$ Earth masses. O'Brien et al. (2010) showed that, in this situation, the Earth could accrete about 0.5-2% of its mass from these objects, enough to supply a few times the current amount of water on our planet (assuming that the primitive planetesimals were

5-10% water by mass). Notice that Walsh et al. and O'Brien et al. did not consider primitive planetary embryos in their simulations, so in principle the total amount of primitive material supplied to the Earth could be somewhat larger than the reported estimate. In view of its smaller yield, the Grand Tack scenario avoids the geochemical inconsistencies that arise from the delivery of a too large amount of carbonaceous chondritic material, characteristic of the classical scenario with giant planets on quasi-circular orbits.

A common feature of the classical and Grand Tack scenarios for the asteroidal delivery of water to the terrestrial planets is that the water is accreted *during* the formation of the planets and not in a *late veneer* fashion (i.e. after that the planets have already achieved their final masses). The accretion of water, though, is not uniform throughout the planet accretion history; instead it accelerates towards the end, as shown in Fig.6.

From the geochemical point of view, Albarede (2009) made a strong case that the deficiency of volatiles (and water) in the Earth relative to solar abundances cannot be explained by invoking the notion that the Earth lost most of its volatiles during its geophysical evolution. In fact, if the Earth had accreted volatiles in solar proportion and had subsequently lost them to space, the abundance of elements would be correlated with their molecular mass. Instead, the relative abundance of elements in our planet is grossly correlated with their condensation temperature. This strongly suggests that the planetesimals and embryos from which the Earth inherited most of its mass were volatile poor because they formed in a disk too hot for the condensation of these elements.

Mann et al. (2009) and Wood et al. (2010) found strong evidence that the accretion of moderately volatile elements to the Earth occurred mostly during the late stages of accretion, but before core formation was complete. The evidence against a late veneer of volatiles is illustrated in Fig.7, which shows that the relative abundance of elements in our planet is correlated not only with condensation temperature but also with chemical affinity. Highly siderophile volatile elements are more depleted in the mantle than moderate siderophile elements or lithophile elements with the *same* condensation temperature. This implies that these volatile elements saw the formation of the Earth's core.

Wood et al. (2008) and Rubie et al. (2011) showed that the oxidation state of the Earth increased progressively as the Earth was growing. Schematically, 60–70% (by mass) of the Earth should have accreted from highly-reduced material with the final 30–40% of accreted mass being more oxidized. This is consistent with the possibility that $H_2O$ was added, together with moderately volatile elements, during the final 30–40% of accretion.

All together these results of geochemical models are in good qualitative agreement with the scenario of terrestrial planet accretion emerging from the astrophysical simulations. This gives confidence that, although many issues have yet to be understood, we are approaching a coherent and global view of the terrestrial planet formation history.

## 8. Summary and Future Prospects

Models of the formation of the terrestrial planets have come a long way since the pioneering models of Wetherill and his students, and now take account of the compelling likelihood of significant giant planet migration within our own solar system. That one might consider such an apparently catastrophic model with a degree of confidence is the result of over a decade of observations of giant planets in a rich variety of orbits around other stars, the evidence from their orbital properties of significant dynamical interactions (Juric' and Tremaine 2008), and the detection by microlensing of a sufficient number of free-floating planets (Sumi et al. 2011) to suggest that giant planet migration and interactions, even catastrophic, might be the rule rather than the exception. Applying this line of thinking to our own Solar System leads to solutions for the small size of Mars, the structure of the asteroid belt and the origin of water on the Earth. However, a number of key problems remain, including the correct treatment of accreting vs. nonaccreting impacts, and the geochemistry of the planetesimals from the giant planet region that seeded the Earth with water.

The future observations required to test this picture are multidisciplinary, ranging from in situ studies of comets to astronomical observations of protoplanetary disks around other stars. Starting in our solar system and moving outward, the key studies are

1. Observations of asteroids that better tie meteorite types to their asteroidal parent bodies. For example, the geochemistry of Ceres, to be assessed by Dawn, and sample returns missions from primitive asteroids will give us more confidence in our associations of meteorite classes with specific asteroid types.
2. A more comprehensive inventory of D/H in water in solar system bodies. This is key to understand connections and differences between primitive asteroids and comets which, in turn, are diagnostic for models of implantation of primitive objects into the asteroid belt from a cometary reservoir, such as the Grand Tack. Further observation of D/H in comets are forthcoming. Also a better understanding of the D/H values in reservoirs of water on Mars would be helpful (Lunine 2003).
3. Determination of the size of heavy element cores in Jupiter and Saturn, and of the oxygen abundances in each, would constrain the composition of the planetesimals formed those planets. This would indirectly provide indications of where Jupiter and Saturn formed and hence more tests for the Grand Tack model. The mission Juno will obtain this information at Jupiter and the Cassini proximal orbits will help provide constraints for Saturn; but obtaining oxygen in Saturn's atmosphere (via water) will be difficult from anything other than a microwave sounder or deep probe.
4. A more complete census of the orbital properties of extrasolar giant planets, and of the occurrence of free-floating giant planets (the latter from microlensing surveys) would better constrain giant planet formation models and the extent to which dynamical interactions are common after formation.
5. High resolution observations of disks from gaseous to transitional, from next generation facilities such as ALMA and JWST can provide information on the timing and extent of giant planet migration.

# FIGURE CAPTIONS

**Fig.1.** An illustration of the process of runaway growth. Each panel represents a snapshot of the system at a different time. The coordinates represent the semi major axis *a* and the eccentricity *e* of orbits of the objects in a portion of the disk centered at 1 AU. The size of the dots is proportional to the physical radius of the objects. Initially, the system is made of a planetesimal population, in which two objects are 2 times more massive than the others. These objects accrete planetesimals very fast, increasing exponentially their mass ratio relative to the individual planetesimals, until they become planetary embryos. Notice how the eccentricity of the planetary embryos remain low, while the eccentricities of the planetesimals are excited with time. The embryos also separate from each other as they grow. At the end, the embryos have grown by a factor 200, whereas the mean mass of the planetesimals has grown only by a factor of 2. From Kokubo and Ida, 1998.

**Fig. 2.** The growth of terrestrial planets from a disk of planetary embryos. Each panel shows the semi-major axis and eccentricity of the bodies in the system at a given time, reported on top of the panel. Embryos and protoplanets are represented with filled dots, whose size is proportional to the cubic root of their mass. Planetesimals are represented by crosses. The big dot at the right-hand side of each panel represents Jupiter, not at scale with respect to the embryos. A system of 3 terrestrial planets, the most massive of which has approximately an Earth mass, is eventually formed inside of 2 AU, whereas only a small fraction of the original planetesimal population survives within the asteroid belt boundaries (sketched with dashed curves). From O'Brien et al. (2006).

**Fig. 3.** The orbits of embryos (green full dots) and planetesimals (red dots) at the end of the inward-then-outward migration of Jupiter as modeled in "the Grand Tack", when the gas is fully removed. The dash curve in the right bottom corner marks the inner boundary of the asteroid belt. From this state, the system evolves naturally in a timescale of a few $10^7$ y into two Earth-mass planets at ~0.7 and 1 AU and a small Mars at 1.5 AU. (see Fig. 4).

**Fig. 4.** The mass distribution of the synthetic terrestrial planets produced in the Walsh et al. (2011) simulations. The open symbols represent the planets produced in different runs starting from different initial conditions. The horizontal lines denote the perihelion-aphelion excursion of the planets on their eccentric final orbits.The black squares show the real planets of the solar system. The large mass ratio between the Earth and Mars is statistically reproduced.

**Fig. 5.** CI and CM meteorites are the most rich in water; water amounts to about 5 to 10% of their total mass (Robert and Epstein, 1982; Kerridge 1985). They are expected to come from C-type asteroids, predominantly in the asteroid belt and possibly accreted even further out (Walsh et al., 2011). Water in ordinary chondrites amounts for only 0.1% of the total weight (Robertet al. 1977; Robert et al. 1979; McNaughton et al. 1981), or a few times as much (Jarosewich 1966); they are spectroscopically linked to S-type asteroids, predominant between 2 and 2.5 AU. Finally, enstatite chondrites are very dry, with only 0.01% of their total mass in water (ref.); they are expected to come from E-type asteroids, which dominate the Hungaria region in the very inner asteroid belt at 1.8 AU.

**Fig.6.** The fraction of the total mass of accreted wet material on a terrestrial planet as a function of the planet 's total mass, from a simulation of O'Brien et al. (2010). In this case, 50% of it's water is accreted when the planet is at least 90% of it's final mass. Assuming that the "wet material" is 5% water by mass, consistent with CM carbonaceous chondrites, this planet would have 4 times the amount of crustal water on Earth at the end of this simulation.

**Fig. 7.** The abundance of elements on Earth, as a function of their condensation temperature and chemical affinity. The open symbols show the abundances of elements in the Allende CV3 chondrite and filled dots those in the silicate Earth, relative to CI chondrites, plotted as a function of 50% condensation temperature. Red symbols show those elements in silicate Earth, which are highly siderophile, black symbols are moderately to weakly siderophile and blue symbols are lithophile. From Wood et al., (2010).

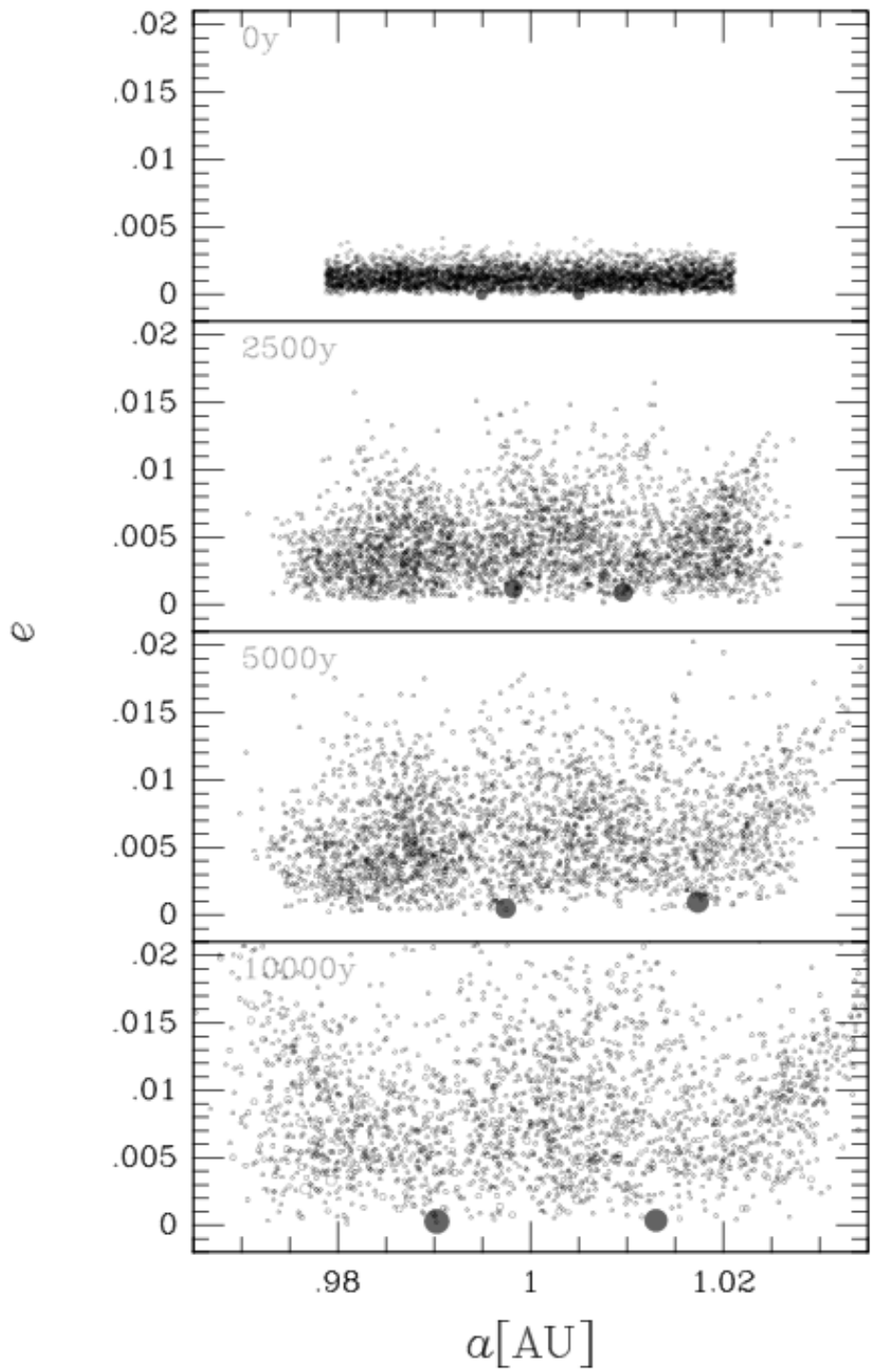

Figure 1

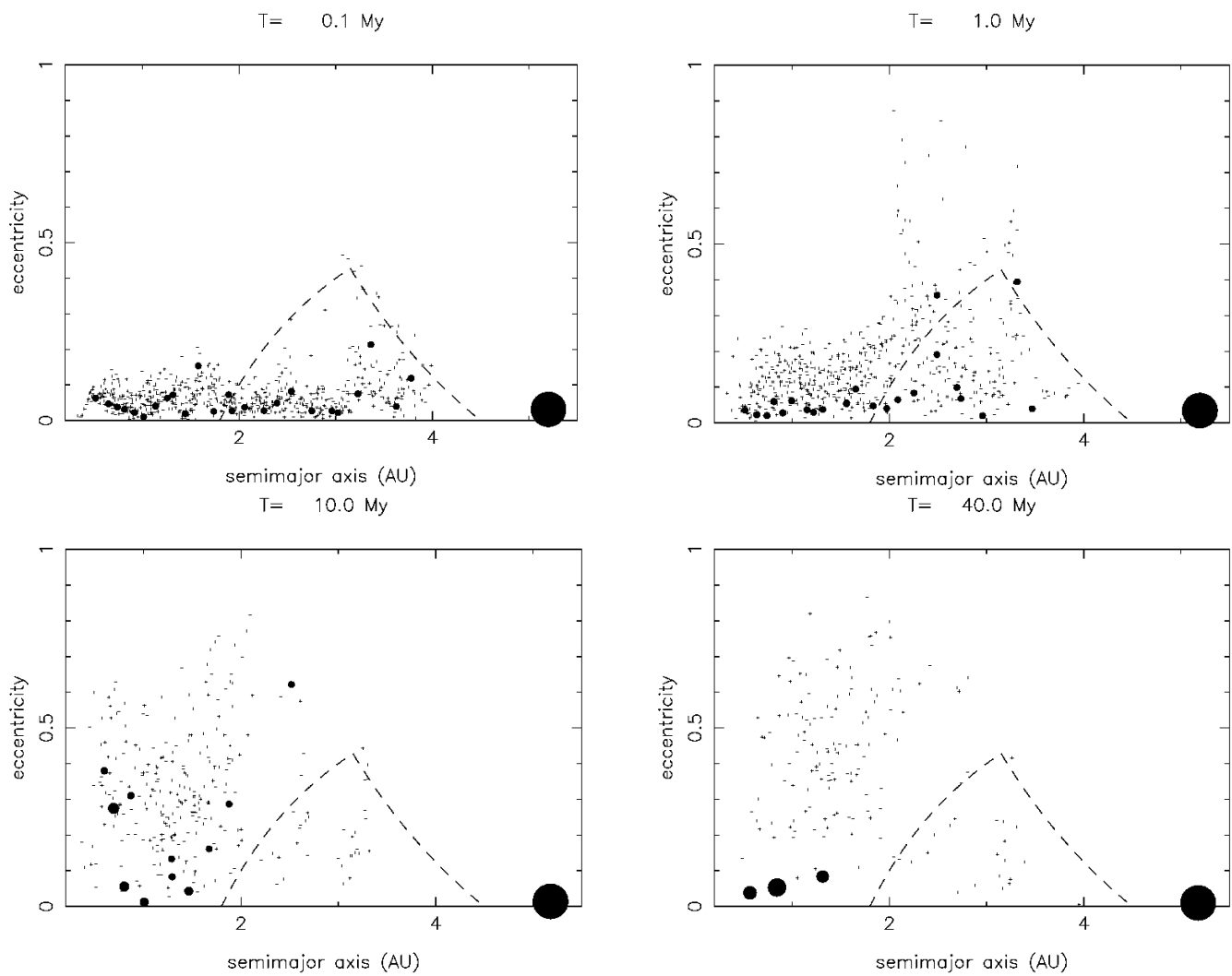

Figure 2

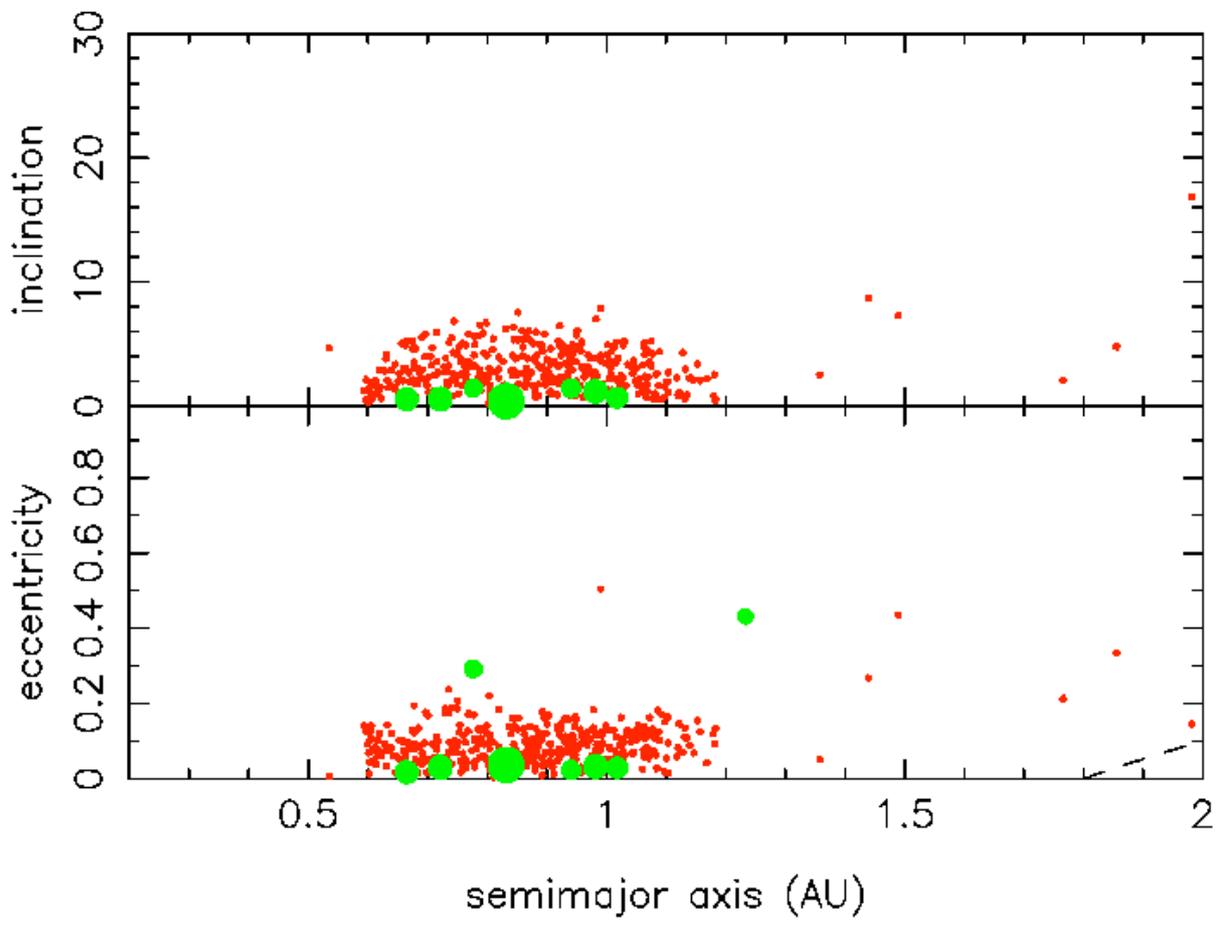

Figure 3

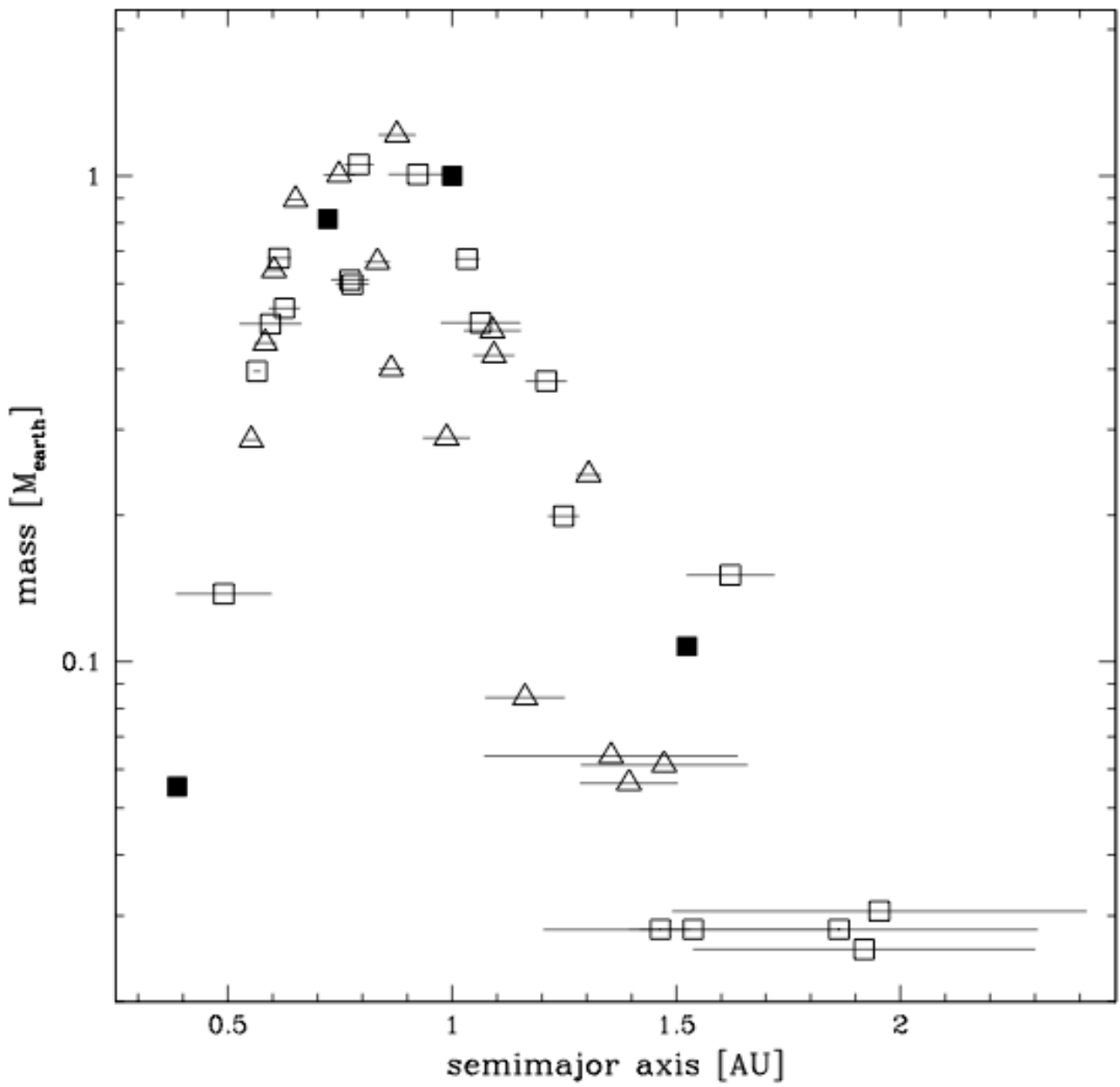

Figure 4

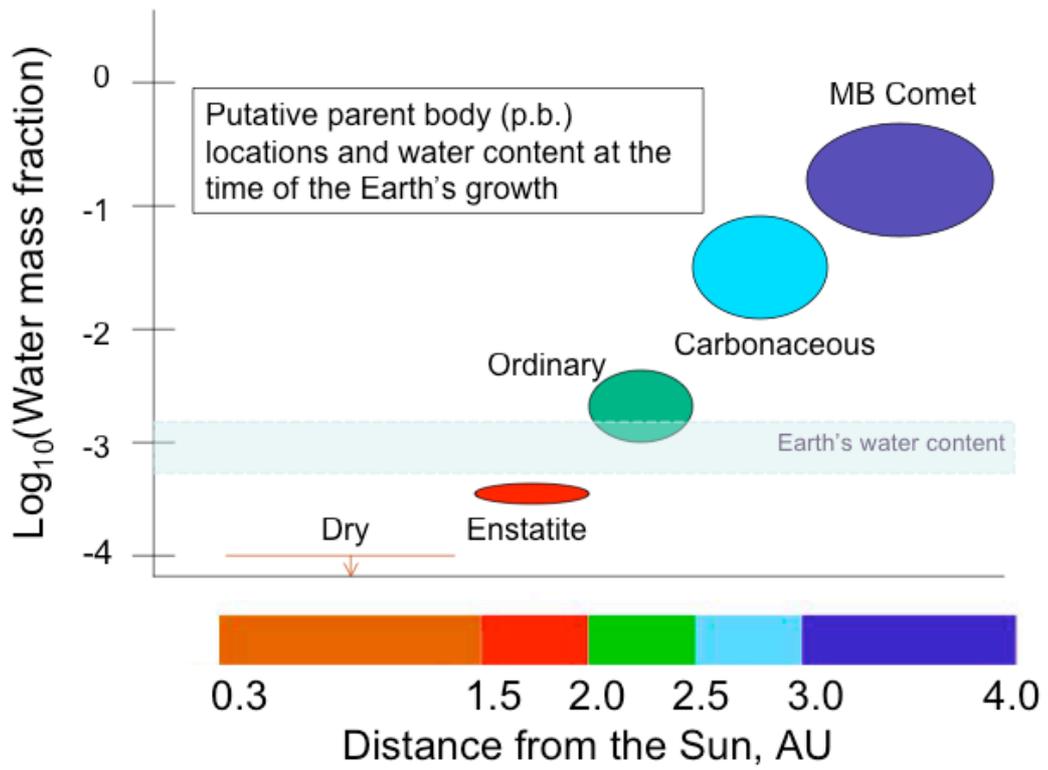

Figure 5

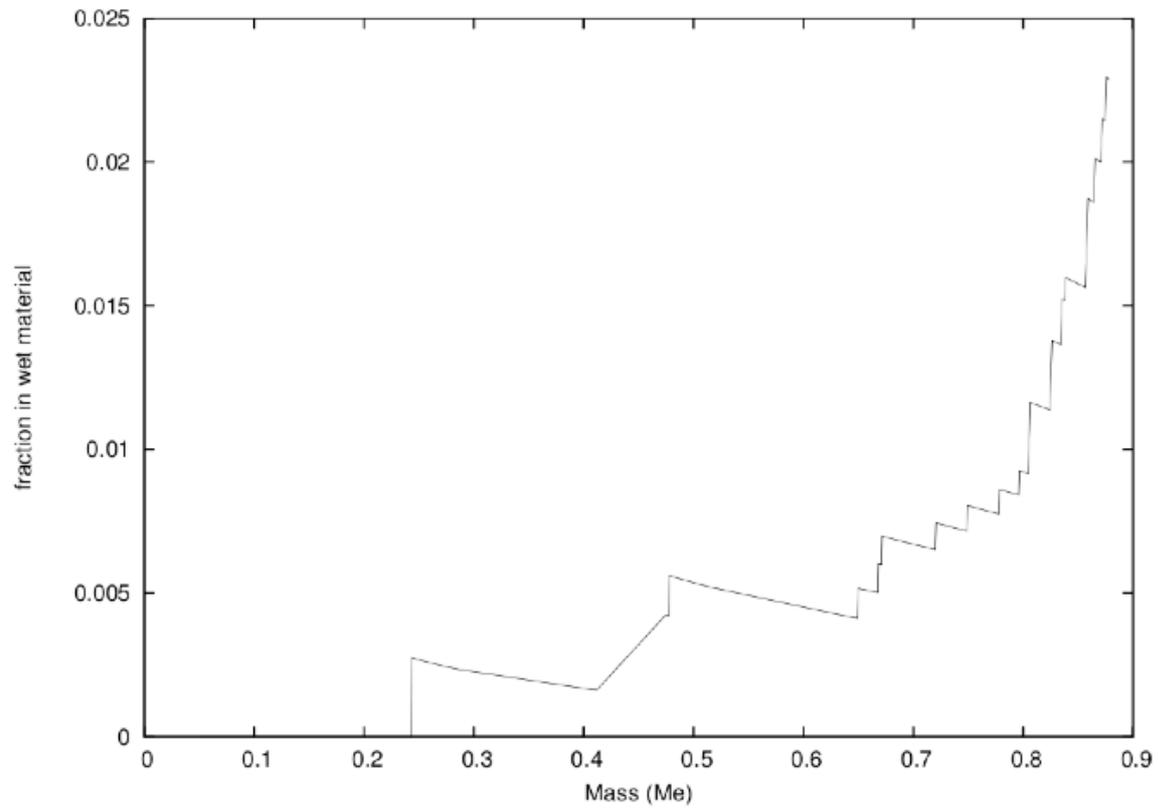

Figure 6

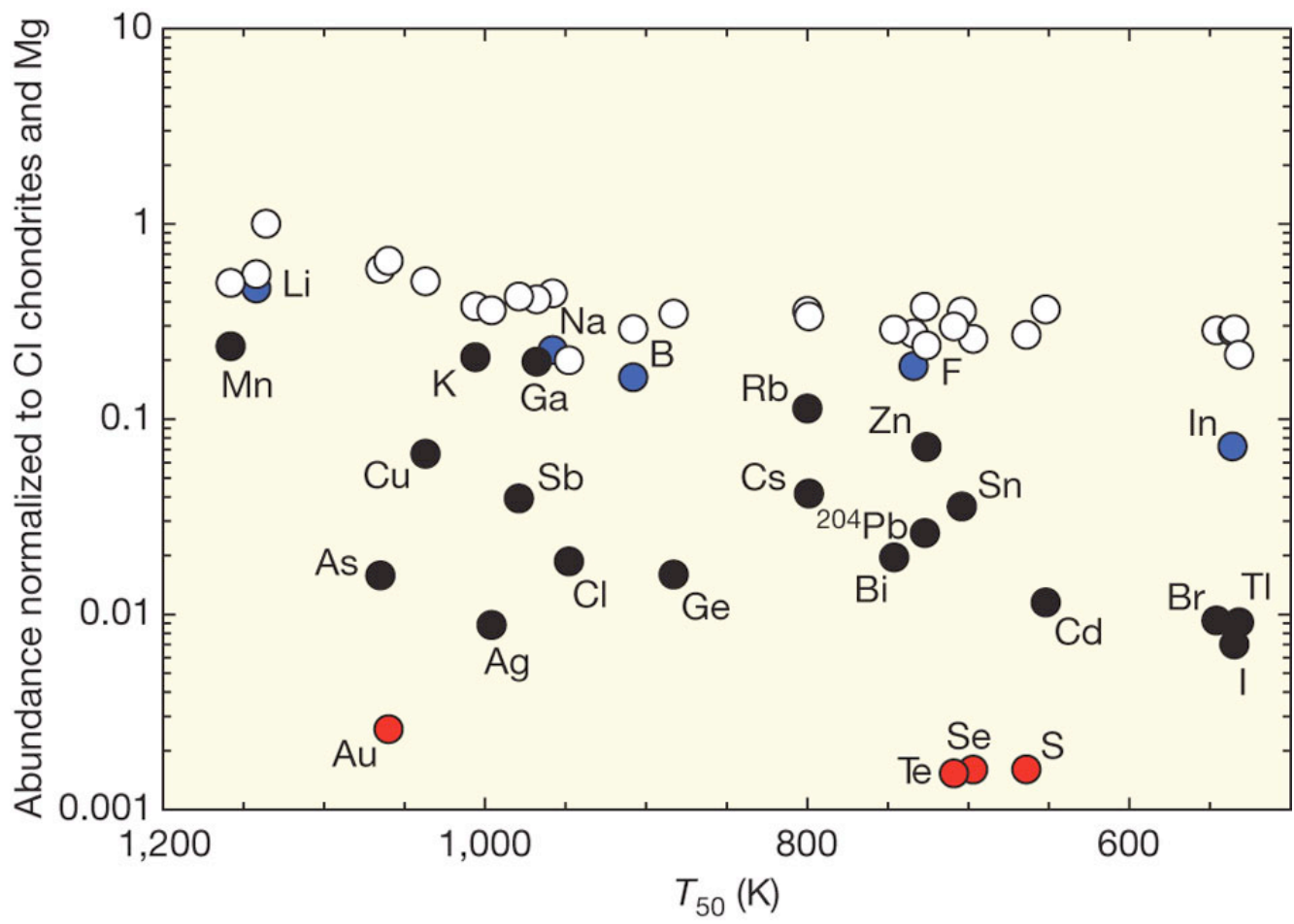

Figure 7